\documentstyle[12pt]{article}

\begin{document}

\author{C.\ Bizdadea\thanks{%
e-mail address: bizdadea@hotmail.com} and S. O. Saliu\thanks{%
e-mail addresses: osaliu@central.ucv.ro or odile\_saliu@hotmail.com} \\
Department of Physics, University of Craiova\\
13 A. I.\ Cuza Str., Craiova R-1100, Romania}
\title{Irreducible Antifield-BRST Approach to Reducible Gauge Theories }
\maketitle

\begin{abstract}
An irreducible antifield BRST quantization method for reducible gauge
theories is proposed. The general formalism is illustrated in the case of
the Freedman-Townsend model.

\end{abstract}

\section{Introduction}

There are two main approaches to the quantization of gauge theories with
open algebras, both related to the BRST symmetry. The first one is based on
the Hamiltonian formalism \cite{1}--\cite{6}, while the second one relies on
the Lagrangian formulation \cite{6}--\cite{11}. Both methods can be applied
to irreducible, as well as to reducible gauge theories. In the irreducible
case the ghosts can be regarded as one-forms dual to the vector fields
associated with the gauge transformations. In the reducible situation this
interpretation fails, being necessary to add ghosts of ghosts together with
their antifields. The ghosts of ghosts are required in order to accommodate
the reducibility relations to the cohomology of the (a model of)
longitudinal exterior differential along the gauge orbits \cite{6}, while
their corresponding antifields ensure the acyclicity of the Koszul-Tate
operator at non-vanishing antighost numbers.

In this paper we propose an irreducible BRST approach to the quantization of
on-shell reducible Lagrangian gauge theories. In consequence, the ghosts of
ghosts and their antifields are absent. Our treatment mainly focuses on: (i)
transforming the initial redundant gauge theory into an irreducible one in a
manner that allows the substitution of the BRST quantization of the
reducible system with that of the irreducible theory, and (ii) quantizing
the irreducible theory along the antifield-BRST ideas. We mention that the
idea of replacing a reducible system by an equivalent irreducible one
appeared for the first time in the Hamiltonian context \cite{6}, \cite{6B}
and was developed recently in the case of the quantization of Hamiltonian
systems with off-shell reducible first-class constraints \cite{Bizd}.

Our paper is structured in five sections. In Section 2 we start with an $L$%
-stage reducible theory, and derive an irreducible system by means of
constructing an irreducible Koszul-Tate differential associated with the
original reducible one. The irreducible Koszul-Tate complex is obtained by
requiring that all the antighost number two co-cycles become trivial under
an appropriate redefinition of the antighost number two antifields. This
request implies the enlargement of both field and antifield spectra. Section
3 focuses on the derivation of the irreducible BRST symmetry corresponding
to the irreducible theory inferred in Section 2, emphasizing that we can
replace the antifield BRST quantization of the irreducible theory by that of
the irreducible system. In Section 4 we illustrate our procedure in the case
of the Freedman-Townsend model. Section 5 ends the paper with some
conclusions.

\section{Derivation of the irreducible theory}

\subsection{The problem}

Our starting point is the gauge invariant Lagrangian action 
\begin{equation}
\label{17}S_0\left[ \Phi ^{\alpha _0}\right] =\int d^Dx{\cal L}_0\left( \Phi
^{\alpha _0},\partial _\mu \Phi ^{\alpha _0},\ldots ,\partial _{\mu
_1}\ldots \partial _{\mu _l}\Phi ^{\alpha _0}\right) , 
\end{equation}
subject to the gauge transformations 
\begin{equation}
\label{18}\delta _\epsilon \Phi ^{\alpha _0}=Z_{\;\;\alpha _1}^{\alpha
_0}\epsilon ^{\alpha _1},\;\alpha _0=1,\ldots ,M_0,\;\alpha _1=1,\ldots
,M_1, 
\end{equation}
which are assumed to be $L$-stage reducible 
\begin{equation}
\label{19}Z_{\;\;\alpha _1}^{\alpha _0}Z_{\;\;\alpha _2}^{\alpha
_1}=C_{\alpha _2}^{\alpha _0\beta _0}\frac{\delta S_0}{\delta \Phi ^{\beta
_0}},\;\alpha _2=1,\ldots ,M_2, 
\end{equation}
\begin{equation}
\label{20}Z_{\;\;\alpha _2}^{\alpha _1}Z_{\;\;\alpha _3}^{\alpha
_2}=C_{\alpha _3}^{\alpha _1\beta _0}\frac{\delta S_0}{\delta \Phi ^{\beta
_0}},\;\alpha _3=1,\ldots ,M_3, 
\end{equation}
$$
\vdots 
$$
\begin{equation}
\label{22}Z_{\;\;\alpha _L}^{\alpha _{L-1}}Z_{\;\;\alpha _{L+1}}^{\alpha
_L}=C_{\alpha _{L+1}}^{\alpha _{L-1}\beta _0}\frac{\delta S_0}{\delta \Phi
^{\beta _0}},\;\alpha _{L+1}=1,\ldots ,M_{L+1}, 
\end{equation}
where $\delta S_0/\delta \Phi ^{\beta _0}=0$ stand for the fields equations.
For the sake of notational simplicity we take the fields to be bosonic. The
subsequent discussion can be straightforwardly extended to fermions modulo
the introduction of some appropriate phase factors.

The reducible BRST symmetry corresponding to the above reducible theory, $%
s_R=\delta _R+\sigma _R+\cdots $, contains two basic differentials. The
first one, $\delta _R$, named the Koszul-Tate differential, realizes an
homological resolution of smooth functions defined on the stationary surface
of field equations, while the second one, $\sigma _R$, represents a model of
longitudinal derivative along the gauge orbits and accounts for the gauge
invariances. For first-stage reducible theories, the construction of $\delta
_R$ requires the introduction of the antifields $\Phi _{\alpha _0}^{*}$, $%
\eta _{\alpha _1}^{*}$ and $C_{\alpha _2}^{*}$, with the Grassmann parities (%
$\varepsilon $) and antighost numbers ($antigh$) given by 
\begin{equation}
\label{i2}\varepsilon \left( \Phi _{\alpha _0}^{*}\right) =1,\;\varepsilon
\left( \eta _{\alpha _1}^{*}\right) =0,\;\varepsilon \left( C_{\alpha
_2}^{*}\right) =1, 
\end{equation}
\begin{equation}
\label{i3}antigh\left( \Phi _{\alpha _0}^{*}\right) =1,\;antigh\left( \eta
_{\alpha _1}^{*}\right) =2,\;antigh\left( C_{\alpha _2}^{*}\right) =3. 
\end{equation}
The standard definitions of $\delta _R$ read as 
\begin{equation}
\label{i4}\delta _R\Phi ^{\alpha _0}=0,\;\delta _R\Phi _{\alpha _0}^{*}=-%
\frac{\delta S_0}{\delta \Phi ^{\alpha _0}}, 
\end{equation}
\begin{equation}
\label{i5}\delta _R\eta _{\alpha _1}^{*}=Z_{\;\;\alpha _1}^{\alpha _0}\Phi
_{\alpha _0}^{*}, 
\end{equation}
\begin{equation}
\label{i6}\delta _RC_{\alpha _2}^{*}=-Z_{\;\;\alpha _2}^{\alpha _1}\eta
_{\alpha _1}^{*}-\frac 12C_{\alpha _2}^{\alpha _0\beta _0}\Phi _{\alpha
_0}^{*}\Phi _{\beta _0}^{*}. 
\end{equation}
The antifields $C_{\alpha _2}^{*}$ are necessary in order to kill the
antighost number two non trivial co-cycles 
\begin{equation}
\label{i7}\nu _{\alpha _2}=Z_{\;\;\alpha _2}^{\alpha _1}\eta _{\alpha
_1}^{*}+\frac 12C_{\alpha _2}^{\alpha _0\beta _0}\Phi _{\alpha _0}^{*}\Phi
_{\beta _0}^{*}, 
\end{equation}
resulting from (\ref{i5}) via the reducibility relations (\ref{19}). In the
case of two-stage reducible theories, apart from the above antifield
spectrum, one should add the antifields $C_{\alpha _3}^{*}$, with $%
\varepsilon \left( C_{\alpha _3}^{*}\right) =0$, $antigh\left( C_{\alpha
_3}^{*}\right) =4$, in order to kill the existing antighost number three
co-cycles yielded by (\ref{i6}) if one takes into account the reducibility
relations. In general, for an $L$-stage reducible system the antifield
spectrum will contain the variables $\Phi _{\alpha _0}^{*}$, $\eta _{\alpha
_1}^{*}$ and $\left( C_{\alpha _k}^{*}\right) _{k=2,\cdots ,L+1}$, where $%
\varepsilon \left( C_{\alpha _k}^{*}\right) =k+1\;{\rm mod}\;2$, $%
antigh\left( C_{\alpha _k}^{*}\right) =k+1$, that are introduced in order to
prevent the appearance of any non trivial co-cycle at positive antighost
numbers.

The problem to be investigated in this section is the derivation of an
irreducible theory associated with a starting $L$-stage reducible gauge
system. In this light, our main idea is to redefine the antifields $\eta
_{\alpha _1}^{*}$ in such a way that all the non trivial co-cycles (\ref{i7}%
) become trivial. The triviality of these co-cycles further implies that the
antifields $\left( C_{\alpha _k}^{*}\right) _{k=2,\cdots ,L+1}$ are no
longer necessary as there are also no non trivial co-cycles at antighost
numbers greater that two. The implementation of this idea leads to an
irreducible gauge theory that possesses the same physical observables like
the original reducible one. In order to clarify our irreducible mechanism,
we gradually investigate the cases $L=1,2$, and then generalize the results
to an arbitrary $L$.

\subsection{The case $L=1$}

Here we start with the relations (\ref{i4}--\ref{i6}) and the reducibility
relations (\ref{19}). In the light of the idea exposed above, we redefine
the antifields $\eta _{\alpha _1}^{*}$ as 
\begin{equation}
\label{i10}\eta _{\alpha _1}^{*}\rightarrow \tilde \eta _{\alpha
_1}^{*}=\eta _{\alpha _1}^{*}-Z_{\;\;_{\beta _2}}^{\beta _1}\bar
D_{\;\;\alpha _2}^{\beta _2}A_{\alpha _1}^{\;\;\alpha _2}\eta _{\beta
_1}^{*}-\frac 12C_{\beta _2}^{\alpha _0\beta _0}\bar D_{\;\;\alpha
_2}^{\beta _2}A_{\alpha _1}^{\;\;\alpha _2}\Phi _{\alpha _0}^{*}\Phi _{\beta
_0}^{*}, 
\end{equation}
where $\bar D_{\;\;\alpha _2}^{\beta _2}$ is the inverse of $D_{\;\;\alpha
_2}^{\beta _2}=Z_{\;\;\alpha _2}^{\alpha _1}A_{\alpha _1}^{\;\;\beta _2}$
and $A_{\alpha _1}^{\;\;\beta _2}$ are some functions that may involve the
fields $\Phi ^{\alpha _0}$, taken such that $rank\left( D_{\;\;\alpha
_2}^{\beta _2}\right) =M_2$. The next step is to replace (\ref{i5}) with 
\begin{equation}
\label{i13}\delta \tilde \eta _{\alpha _1}^{*}=Z_{\;\;\alpha _1}^{\alpha
_0}\Phi _{\alpha _0}^{*}. 
\end{equation}
The relations (\ref{i13}) lead to some co-cycles of the type (\ref{i7}),
i.e., 
\begin{equation}
\label{i14}\tilde \nu _{\alpha _2}=Z_{\;\;\alpha _2}^{\alpha _1}\tilde \eta
_{\alpha _1}^{*}+\frac 12C_{\alpha _2}^{\alpha _0\beta _0}\Phi _{\alpha
_0}^{*}\Phi _{\beta _0}^{*}, 
\end{equation}
that are trivial by virtue of (\ref{i10}). Indeed, from (\ref{i10}) we find 
\begin{equation}
\label{i15}Z_{\;\;\alpha _2}^{\alpha _1}\tilde \eta _{\alpha _1}^{*}=-\frac
12C_{\alpha _2}^{\alpha _0\beta _0}\Phi _{\alpha _0}^{*}\Phi _{\beta
_0}^{*}, 
\end{equation}
hence $\tilde \nu _{\alpha _2}\equiv 0$. In consequence, the relations (\ref
{i13}) do not imply any non trivial co-cycles at antighost number two, so
the antifields $C_{\alpha _2}^{*}$ are no longer necessary. Thus, formula (%
\ref{i13}) helps us at deriving an irreducible theory. This is the reason
for changing the notation $\delta _R$ into $\delta $ in (\ref{i13}). In
order to infer the irreducible gauge transformations corresponding to the
irreducible theory we introduce the fields $\Phi ^{\alpha _2}$ and require
that their antifields, denoted by $\Phi _{\alpha _2}^{*}$, are the non
vanishing solutions to the equations 
\begin{equation}
\label{i16}D_{\;\;\beta _2}^{\alpha _2}\Phi _{\alpha _2}^{*}=\delta \left(
Z_{\;\;\beta _2}^{\alpha _1}\eta _{\alpha _1}^{*}+\frac 12C_{\beta
_2}^{\alpha _0\beta _0}\Phi _{\alpha _0}^{*}\Phi _{\beta _0}^{*}\right) . 
\end{equation}
The $\Phi _{\alpha _2}^{*}$'s are fermionic and possess antighost number
one. Due to the invertibility of $D_{\;\;\beta _2}^{\alpha _2}$, the non
vanishing solutions for $\Phi _{\alpha _2}^{*}$ enforce the irreducibility
because the equations (\ref{i16}) possess non vanishing solutions if and
only if 
\begin{equation}
\label{i17}\delta \left( Z_{\;\;\beta _2}^{\alpha _1}\eta _{\alpha
_1}^{*}+\frac 12C_{\beta _2}^{\alpha _0\beta _0}\Phi _{\alpha _0}^{*}\Phi
_{\beta _0}^{*}\right) \neq 0, 
\end{equation}
hence if and only if (\ref{i7}) are not co-cycles. In the meantime, the
invertibility of $D_{\;\;\beta _2}^{\alpha _2}$ emphasizes via (\ref{i16})
that the antifields $\Phi _{\alpha _2}^{*}$ are $\delta $-exact, which then
ensures by virtue of the nilpotency of $\delta $ that 
\begin{equation}
\label{i17a}\delta \Phi _{\alpha _2}^{*}=0. 
\end{equation}
With the help of the relations (\ref{i10}), (\ref{i13}) and (\ref{i16}) we
arrive at 
\begin{equation}
\label{i18}\delta \eta _{\alpha _1}^{*}=Z_{\;\;\alpha _1}^{\alpha _0}\Phi
_{\alpha _0}^{*}+A_{\alpha _1}^{\;\;\alpha _2}\Phi _{\alpha _2}^{*}. 
\end{equation}
By maintaining the definitions from the reducible case 
\begin{equation}
\label{i19}\delta \Phi ^{\alpha _0}=0,\;\delta \Phi _{\alpha _0}^{*}=-\frac{%
\delta S_0}{\delta \Phi ^{\alpha _0}}, 
\end{equation}
and by setting 
\begin{equation}
\label{i20}\delta \Phi ^{\alpha _2}=0, 
\end{equation}
the relations (\ref{i17a}--\ref{i20}) completely define the irreducible
Koszul-Tate complex corresponding to an irreducible theory associated with
the original reducible one. At this point we can deduce the action of the
irreducible theory $\tilde S_0\left[ \Phi ^{\alpha _0},\Phi ^{\alpha
_2}\right] $, as well as its gauge invariances. On the one hand, from the
standard BRST prescription 
\begin{equation}
\label{i21}\delta \Phi _{\alpha _2}^{*}=-\frac{\delta \tilde S_0}{\delta
\Phi ^{\alpha _2}}, 
\end{equation}
compared with (\ref{i17a}), we find that 
\begin{equation}
\label{i22}\tilde S_0\left[ \Phi ^{\alpha _0},\Phi ^{\alpha _2}\right]
=S_0\left[ \Phi ^{\alpha _0}\right] . 
\end{equation}
On the other hand, the formulas (\ref{i18}) lead to the gauge
transformations of the irreducible theory under the form 
\begin{equation}
\label{i23}\delta _\epsilon \Phi ^{\alpha _0}=Z_{\;\;\alpha _1}^{\alpha
_0}\epsilon ^{\alpha _1},\;\delta _\epsilon \Phi ^{\alpha _2}=A_{\alpha
_1}^{\;\;\alpha _2}\epsilon ^{\alpha _1}. 
\end{equation}
Thus, we can conclude that the irreducible theory is based on the original
action (see (\ref{i22})) and the gauge transformations (\ref{i23}). From (%
\ref{i22}) it is clear that the fields $\Phi ^{\alpha _2}$ are purely gauge,
such that the physical observables of the irreducible system coincide with
those of the original reducible theory. The equivalence between the physical
observables represents a desirable feature of our irreducible method, which
can be gained if we set all the antifields corresponding to the new
introduced fields to be $\delta $-closed. On the other hand, as these
antifields should not represent non trivial co-cycles, it is necessary to
construct the theory such that they are also $\delta $-exact. Anticipating a
bit, we remark that for higher order reducible theories it is necessary to
further enlarge the field and antifield spectra in order to enforce the
above discussed $\delta $-exactness.

\subsection{The case $L=2$}

In this situation we start with the definitions of $\delta $ given by (\ref
{i17a}--\ref{i20}). However, we have to account in addition of the second
stage reducibility relations (\ref{20}). On behalf of these supplementary
reducibility relations, we find that the matrix $D_{\;\;\beta _2}^{\alpha
_2} $ is no longer invertible, as it displays some on-shell null vectors,
namely, 
\begin{equation}
\label{i24}D_{\;\;\beta _2}^{\alpha _2}Z_{\;\;\beta _3}^{\beta _2}=A_{\beta
_1}^{\;\;\alpha _2}C_{\beta _3}^{\beta _1\beta _0}\frac{\delta S_0}{\delta
\Phi ^{\beta _0}}\approx 0, 
\end{equation}
where the weak equality `$\approx $' means an equality valid when the field
equations hold. Thus, in the case of two-stage reducible theories we will
consider that $A_{\alpha _1}^{\;\;\alpha _2}$ are chosen such that $%
rank\left( D_{\;\;\beta _2}^{\alpha _2}\right) \approx M_2-M_3$. Multiplying
(\ref{i18}) by $Z_{\;\;\beta _2}^{\alpha _1}$, we obtain 
\begin{equation}
\label{i24a}\delta \left( Z_{\;\;\beta _2}^{\alpha _1}\eta _{\alpha
_1}^{*}+\frac 12C_{\beta _2}^{\alpha _0\beta _0}\Phi _{\alpha _0}^{*}\Phi
_{\beta _0}^{*}\right) =D_{\;\;\beta _2}^{\alpha _2}\Phi _{\alpha _2}^{*}, 
\end{equation}
that together with (\ref{i24}) and (\ref{i17a}--\ref{i20}) lead to the
antighost number two co-cycles 
\begin{equation}
\label{i25}\nu _{\alpha _3}=Z_{\;\;\alpha _3}^{\alpha _2}Z_{\;\;\alpha
_2}^{\alpha _1}\eta _{\alpha _1}^{*}+\frac 12C_{\alpha _2}^{\alpha _0\beta
_0}Z_{\;\;\alpha _3}^{\alpha _2}\Phi _{\alpha _0}^{*}\Phi _{\beta
_0}^{*}+A_{\beta _1}^{\;\;\alpha _2}C_{\alpha _3}^{\beta _1\beta _0}\Phi
_{\alpha _2}^{*}\Phi _{\beta _0}^{*}, 
\end{equation}
which are found trivial, $\nu _{\alpha _3}=\delta \left( -C_{\alpha
_3}^{\beta _1\beta _0}\Phi _{\beta _0}^{*}\eta _{\beta _1}^{*}\right) $, so
there are actually no non trivial co-cycles at antighost number two. In this
way, the only problem to be solved remains the $\delta $-exactness of $\Phi
_{\alpha _2}^{*}$, which will further ensure that there are no non trivial
co-cycles at antighost number one. The equations (\ref{i24}) allow us to
represent $D_{\;\;\beta _2}^{\alpha _2}$ under the form 
\begin{equation}
\label{i27}D_{\;\;\beta _2}^{\alpha _2}=\delta _{\;\;\beta _2}^{\alpha
_2}-Z_{\;\;\alpha _3}^{\alpha _2}\bar D_{\;\;\beta _3}^{\alpha _3}A_{\beta
_2}^{\;\;\beta _3}+A_{\beta _1}^{\;\;\alpha _2}C_{\alpha _3}^{\beta _1\beta
_0}\bar D_{\;\;\beta _3}^{\alpha _3}A_{\beta _2}^{\;\;\beta _3}\frac{\delta
S_0}{\delta \Phi ^{\beta _0}}, 
\end{equation}
where $\bar D_{\;\;\beta _3}^{\alpha _3}$ is the inverse of $D_{\;\;\beta
_3}^{\alpha _3}=Z_{\;\;\beta _3}^{\beta _2}A_{\beta _2}^{\;\;\alpha _3}$ and 
$A_{\beta _2}^{\;\;\alpha _3}$ are some functions that may involve the
fields $\Phi ^{\alpha _0}$, chosen to satisfy $rank\left( D_{\;\;\beta
_3}^{\alpha _3}\right) =M_3$. Inserting (\ref{i27}) in (\ref{i24a}) we
arrive at%
\begin{eqnarray}\label{i30}
& &\delta \left( Z_{\;\;\beta _2}^{\alpha _1}\eta _{\alpha _1}^{*}+\frac
12C_{\beta _2}^{\alpha _0\beta _0}\Phi _{\alpha _0}^{*}\Phi _{\beta
_0}^{*}+A_{\beta _1}^{\;\;\alpha _2}C_{\alpha _3}^{\beta _1\beta _0}\bar
D_{\;\;\beta _3}^{\alpha _3}A_{\beta _2}^{\;\;\beta _3}\Phi _{\alpha
_2}^{*}\Phi _{\beta _0}^{*}\right) =\nonumber \\ 
& &\Phi _{\beta _2}^{*}-Z_{\;\;\alpha _3}^{\alpha _2}\bar
D_{\;\;\beta _3}^{\alpha _3}A_{\beta _2}^{\;\;\beta _3}\Phi _{\alpha
_2}^{*}, 
\end{eqnarray}
which show that $\Phi _{\beta _2}^{*}$ are not $\delta $-exact in the
context of the present antifield spectrum. In order to restore the $\delta $%
-exactness of $\Phi _{\beta _2}^{*}$ we introduce the bosonic antighost
number two antifields $\eta _{\alpha _3}^{*}$ and define 
\begin{equation}
\label{i31}\delta \eta _{\alpha _3}^{*}=Z_{\;\;\alpha _3}^{\alpha _2}\Phi
_{\alpha _2}^{*}. 
\end{equation}
Introducing the definitions (\ref{i31}) in the relations(\ref{i30}) we
deduce that 
\begin{eqnarray}\label{i32}
& &\Phi _{\beta _2}^{*}=
\delta \left( Z_{\;\;\beta _2}^{\alpha _1}\eta _{\alpha
_1}^{*}+\frac 12C_{\beta _2}^{\alpha _0\beta _0}\Phi _{\alpha _0}^{*}\Phi
_{\beta _0}^{*}+\right. \nonumber \\
& &\left. A_{\beta _1}^{\;\;\alpha _2}C_{\alpha _3}^{\beta _1\beta
_0}\bar D_{\;\;\beta _3}^{\alpha _3}A_{\beta _2}^{\;\;\beta _3}\Phi _{\alpha
_2}^{*}\Phi _{\beta _0}^{*}+\bar D_{\;\;\beta _3}^{\alpha _3}A_{\beta
_2}^{\;\;\beta _3}\eta _{\alpha _3}^{*}\right) , 
\end{eqnarray}
which show that $\Phi _{\beta _2}^{*}$ can be made $\delta $-exact.
Replacing (\ref{i32}) in (\ref{i18}) we get that $Z_{\;\;\alpha _1}^{\alpha
_0}\Phi _{\alpha _0}^{*}$ are also trivial co-cycles. In conclusion, in the
case $L=2$ the formulas (\ref{i17a}--\ref{i20}) and (\ref{i31}) completely
define the irreducible Koszul-Tate complex. Thus, the irreducible theory is
based also on action (\ref{i22}), subject to the gauge transformations 
\begin{equation}
\label{i33}\delta _\epsilon \Phi ^{\alpha _0}=Z_{\;\;\alpha _1}^{\alpha
_0}\epsilon ^{\alpha _1},\;\delta _\epsilon \Phi ^{\alpha _2}=A_{\alpha
_1}^{\;\;\alpha _2}\epsilon ^{\alpha _1}+Z_{\;\;\alpha _3}^{\alpha
_2}\epsilon ^{\alpha _3}, 
\end{equation}
where $\epsilon ^{\alpha _3}$ are some additional gauge parameters due to (%
\ref{i31}).

From the above analysis for $L=1,2$ it seems that there appear some problems
linked with locality. Indeed, the matrices $\bar D_{\;\;\beta _2}^{\alpha
_2} $ present in (\ref{i10}) and (\ref{i13}), and also the solution of the
equations (\ref{i16}) are non-local in general. However, the non-locality
involved with (\ref{i16}) compensates in a certain way that from the
relations (\ref{i13}), such that the irreducible gauge transformations (\ref
{i23}) are local. A similar observation can be made with respect to the case 
$L=2$. In conclusion, the non-locality present within the intermediate steps
of the construction of the irreducible Koszul-Tate complex plays no role in
the irreducible theory. Moreover, the non-locality mentioned in the above
brings no contribution when comparing the results inferred within the
irreducible and reducible procedures (see Section 4).

\subsection{Generalization to arbitrary $L$}

At this point we can generalize the previous results to an arbitrary $L$ in
a simple manner. Acting in a way that ensures on the one hand the nilpotency
and acyclicity of the Koszul-Tate differential, and on the other hand its
irreducibility, we enlarge the field and antifield spectra, and construct
the Koszul-Tate complex through 
\begin{equation}
\label{k1}\delta \Phi ^{\alpha _0}=0,\;\delta \Phi ^{\alpha
_{2k}}=0,\;k=1,\cdots ,a, 
\end{equation}
\begin{equation}
\label{k2}\delta \Phi _{\alpha _0}^{*}=-\frac{\delta S_0}{\delta \Phi
^{\alpha _0}},\;\delta \Phi _{\alpha _{2k}}^{*}=0,\;k=1,\cdots ,a, 
\end{equation}
\begin{equation}
\label{k3}\delta \eta _{\alpha _{2k+1}}^{*}=Z_{\;\;\alpha _{2k+1}}^{\alpha
_{2k}}\Phi _{\alpha _{2k}}^{*}+A_{\alpha _{2k+1}}^{\;\;\alpha _{2k+2}}\Phi
_{\alpha _{2k+2}}^{*},\;k=0,\cdots ,b, 
\end{equation}
where the $\Phi ^{\alpha _{2k}}$'s are bosonic with antighost number zero,
the $\Phi _{\alpha _{2k}}^{*}$'s are fermionic, of antighost number one, and
the $\eta _{\alpha _{2k+1}}^{*}$'s are bosonic with antighost number two. In
the above, the notations $a$ and $b$ mean 
\begin{equation}
\label{53}a=\left\{ 
\begin{array}{c}
\frac L2,\; 
{\rm for}\;L\;{\rm even}, \\ \frac{L+1}2,\;{\rm for}\;L\;{\rm odd}, 
\end{array}
\right. \;\;\;b=\left\{ 
\begin{array}{c}
\frac L2,\; 
{\rm for}\;L\;{\rm even}, \\ \frac{L-1}2,\;{\rm for}\;L\;{\rm odd}. 
\end{array}
\right. 
\end{equation}
From (\ref{k1}--\ref{k3}) we get an irreducible theory described by the
action 
\begin{equation}
\label{55}\tilde S_0\left[ \Phi ^{\alpha _0},\left( \Phi ^{\alpha
_{2k}}\right) _{k=1,\cdots ,a}\right] =S_0\left[ \Phi ^{\alpha _0}\right] , 
\end{equation}
subject to the gauge transformations 
\begin{equation}
\label{56}\delta _\epsilon \Phi ^{\alpha _0}=Z_{\;\;\alpha _1}^{\alpha
_0}\epsilon ^{\alpha _1}, 
\end{equation}
\begin{equation}
\label{57}\delta _\epsilon \Phi ^{\alpha _2}=A_{\alpha _1}^{\;\;\alpha
_2}\epsilon ^{\alpha _1}+Z_{\;\;\alpha _3}^{\alpha _2}\epsilon ^{\alpha _3}, 
\end{equation}
$$
\vdots 
$$
\begin{equation}
\label{58}\delta _\epsilon \Phi ^{\alpha _{2k}}=A_{\alpha
_{2k-1}}^{\;\;\alpha _{2k}}\epsilon ^{\alpha _{2k-1}}+Z_{\;\;\alpha
_{2k+1}}^{\alpha _{2k}}\epsilon ^{\alpha _{2k+1}}, 
\end{equation}
$$
\vdots 
$$
\begin{equation}
\label{59}\delta _\epsilon \Phi ^{\alpha _{2a}}=\left\{ 
\begin{array}{l}
A_{\alpha _{L-1}}^{\;\;\alpha _L}\epsilon ^{\alpha _{L-1}}+Z_{\;\;\alpha
_{L+1}}^{\alpha _L}\epsilon ^{\alpha _{L+1}},\; 
{\rm for}\;L\;{\rm even}, \\ A_{\alpha _L}^{\;\;\alpha _{L+1}}\epsilon
^{\alpha _L},\;\;{\rm for}\;L\;{\rm odd}. 
\end{array}
\right. 
\end{equation}
The functions $A_{\alpha _{2k-1}}^{\;\;\alpha _{2k}}$ may depend on the
fields $\Phi ^{\alpha _0}$ and are chosen to satisfy 
\begin{equation}
\label{r}rank\left( D_{\;\;\alpha _k}^{\beta _k}\right) \approx
\sum\limits_{i=k}^{L+1}\left( -\right) ^{k+i}M_i,\;k=1,\cdots ,L, 
\end{equation}
\begin{equation}
\label{ra}rank\left( D_{\;\;\alpha _{L+1}}^{\beta _{L+1}}\right) =M_{L+1}, 
\end{equation}
where $D_{\;\;\alpha _k}^{\beta _k}=A_{\alpha _{k-1}}^{\;\;\beta
_k}Z_{\;\;\alpha _k}^{\alpha _{k-1}}$. We remark that the choice of the
functions $A_{\alpha _{k-1}}^{\;\;\alpha _k}$ is not unique. Moreover, for a
definite choice of $A_{\alpha _{k-1}}^{\;\;\alpha _k}$, the relations (\ref
{r}--\ref{ra}) are unaffected if we modify the functions $A_{\alpha
_{k-1}}^{\;\;\alpha _k}$ like 
\begin{equation}
\label{r1}A_{\alpha _{k-1}}^{\;\;\alpha _k}\rightarrow A_{\alpha
_{k-1}}^{\;\;\alpha _k}+\mu _{\;\;\beta _{k-2}}^{\alpha _k}Z_{\;\;\alpha
_{k-1}}^{\beta _{k-2}}, 
\end{equation}
so these functions carry some ambiguities. It is known that the reducibility
functions $Z_{\;\;\alpha _k}^{\alpha _{k-1}}$ also display some ambiguities 
\cite{6}. Throughout the paper we use the conventions $f^{\alpha _k}=0$ if $%
k<0$ or $k>L+1$.

\section{The irreducible BRST symmetry for reducible gauge theories}

The derivation of the irreducible Koszul-Tate complex from the above section
suggests the possibility to construct an irreducible BRST symmetry
associated with the reducible one. This is why in this section we point out
the derivation of the irreducible BRST symmetry corresponding to the
irreducible theory derived within the previous section and show that we can
replace the BRST quantization of the original reducible system by that of
the irreducible theory. In view of this, we remark that by organizing the
fields $\left( \Phi ^{\alpha _0},\Phi ^{\alpha _{2k}}\right) $, as well as
the gauge parameters $\left( \epsilon ^{\alpha _1},\epsilon ^{\alpha
_{2k+1}}\right) $, into some column vectors $\Phi ^{A_0}$, respectively, $%
\epsilon ^{A_1}$, the gauge transformations (\ref{56}--\ref{59}) can be
written in a condensed form as $\delta _\epsilon \Phi
^{A_0}=Z_{\;\;A_1}^{A_0}\epsilon ^{A_1}$, where $Z_{\;\;A_1}^{A_0}$ is the
appropriate matrix of the gauge generators from (\ref{56}--\ref{59})
(including $A_{\alpha _{k-1}}^{\;\;\alpha _k}$ and $Z_{\;\;\alpha
_k}^{\alpha _{k-1}}$). An essential requirement that must be satisfied by
the new generators $Z_{\;\;A_1}^{A_0}$ is their completeness, i.e., 
\begin{equation}
\label{68}Z_{\;\;A_1}^{B_0}\frac{\delta Z_{\;\;B_1}^{A_0}}{\delta \Phi ^{B_0}%
}-Z_{\;\;B_1}^{B_0}\frac{\delta Z_{\;\;A_1}^{A_0}}{\delta \Phi ^{B_0}}%
\approx C_{\;\;A_1B_1}^{C_1}Z_{\;\;C_1}^{A_0}. 
\end{equation}
As the completeness of the gauge generators depends in general on the choice
of $A_{\alpha _{k-1}}^{\;\;\alpha _k}$ and also on the reducibility
functions of the original theory, in the sequel we consider only those
theories for which (\ref{68}) hold.

In order to build the irreducible antifield BRST symmetry it is necessary to
construct the irreducible Koszul-Tate differential and the irreducible
longitudinal exterior derivative along the gauge orbits. The Koszul-Tate
differential was constructed in Section 2 (see (\ref{k1}--\ref{k3})). The
construction of the longitudinal exterior differential along the gauge
orbits, $D$, follows the general irreducible BRST line \cite{6}. By
introducing the minimal ghosts 
\begin{equation}
\label{90}\eta ^{A_1}=\left( 
\begin{array}{c}
\eta ^{\alpha _1} \\ 
\eta ^{\alpha _{2k+1}} 
\end{array}
\right) , 
\end{equation}
of pure ghost number one, the definitions 
\begin{equation}
\label{91}D\Phi ^{A_0}=Z_{\;\;A_1}^{A_0}\eta ^{A_1},\;D\eta ^{A_1}=\frac
12C_{\;\;B_1C_1}^{A_1}\eta ^{B_1}\eta ^{C_1}, 
\end{equation}
together with (\ref{68}) ensure the weak nilpotency of $D$ without adding
any ghosts of ghosts. Under these circumstances, the homological
perturbation theory \cite{12}--\cite{15} guarantees the existence of the
irreducible BRST symmetry, $s_I$.

In the sequel we show that it is permissible to substitute the BRST
quantization of the reducible theory by that of the irreducible system
derived previously. It is obvious that the two theories possess the same
classical observables as the fields $\left( \Phi ^{\alpha _{2k}}\right)
_{k=1,\cdots ,a}$ do not appear effectively in the action of the irreducible
system, such that they are purely gauge variables, and, in consequence, the
observables of the irreducible theory do not depend on them, such that $%
\frac{\delta F}{\delta \Phi ^{\alpha _{2k}}}\approx 0$. Thus, the
observables corresponding to the irreducible system, $F$, involve only the
fields $\Phi ^{\alpha _0}$ and should satisfy only the equations $\frac{%
\delta F}{\delta \Phi ^{\alpha _0}}Z_{\;\;\alpha _1}^{\alpha _0}\approx 0$,
which are nothing but the equations that must be checked by the observables
of the reducible theory. As the observables of the irreducible and reducible
theories coincide, it follows that the zeroth order cohomological groups of
the irreducible and reducible BRST operators are isomorphic, $H^0\left(
s_I\right) =H^0\left( s_R\right) $. Thus, the irreducible and reducible
theories are equivalent from the BRST point of view, i.e., from the point of
view of the fundamental equations underlying this formalism, $s^2=0$, $%
H^0\left( s\right) =\left\{ {\rm physical\;observables}\right\} $. All these
considerations lead to the conclusion that we can replace the BRST
quantization of the reducible theory by that of the irreducible system
derived previously.

With all the above ingredients at hand, the BRST quantization of the
irreducible theory goes along the standard manner. If one defines the
canonical action of $s_I$ through $s_IF=\left( F,S_I\right) $, with $\left(
,\right) $ the antibracket and $S_I$ the canonical generator of the
irreducible BRST symmetry, the nilpotency of $s_I$ is expressed by means of
the master equation 
\begin{equation}
\label{93}\left( S_I,S_I\right) =0.
\end{equation}
The existence of the solution to the master equation is guaranteed via the
acyclicity of the Koszul-Tate operator at positive antighost numbers. In
order to solve the master equation we take $S_I=\sum\limits_{k=0}^\infty 
\stackrel{(k)}{S}$, with $antigh\left( \stackrel{(k)}{S}\right) =k$,\ $%
gh\left( \stackrel{(k)}{S}\right) =0$ and approach the master equation (\ref
{93}) antighost by antighost level, requiring at the same time the boundary
conditions%
\begin{eqnarray}\label{95}
\stackrel{(0)}{S}&=&S_0,\;\stackrel{(1)}{S}=
\Phi _{\alpha _0}^{*}Z_{\;\;\alpha
_1}^{\alpha _0}\eta ^{\alpha _1}+\nonumber \\ 
& &\sum\limits_{k=1}^a\Phi _{\alpha _{2k}}^{*}\left( A_{\alpha
_{2k-1}}^{\;\;\alpha _{2k}}\eta ^{\alpha _{2k-1}}+Z_{\;\;\alpha
_{2k+1}}^{\alpha _{2k}}\eta ^{\alpha _{2k+1}}\right) .
\end{eqnarray}
The ambiguities signalized at the end of Section 2 in connection with the
functions $A_{\alpha _{k-1}}^{\;\;\alpha _k}$ and $Z_{\;\;\alpha _k}^{\alpha
_{k-1}}$ induce some ambiguities at the level of the solution to the master
equation, $S_I$. The ambiguity in $S_I$ is completely exhausted by the
possibility to perform a canonical transformation in the antibracket \cite{6}%
, so the solution is unique up to such a canonical transformation. In order
to fix the gauge, we add the non-minimal variables $\left( B^{\alpha
_{2k+1}},B_{\alpha _{2k+1}}^{*}\right) $ and\ $\left( \bar \eta ^{\alpha
_{2k+1}},\bar \eta _{\alpha _{2k+1}}^{*}\right) $, with $k=0,\cdots ,b$,
such that we obtain the non-minimal solution $S=S_I+\sum\limits_{k=0}^b\bar
\eta _{\alpha _{2k+1}}^{*}B^{\alpha _{2k+1}}$. A class of appropriate
gauge-fixing conditions is given by 
\begin{eqnarray}\label{98a}
\chi _{\beta _{2k+1}}&\equiv &Z_{\;\;\beta _{2k+1}}^{\beta _{2k}}f_{\beta
_{2k}}\left( \Phi ^{\beta _{2k}}\right) +\nonumber \\ 
& &A_{\beta _{2k+1}}^{\;\;\beta _{2k+2}}g_{\beta _{2k+2}}\left( \Phi
^{\beta _{2k+2}}\right) =0,
\end{eqnarray}
where 
\begin{equation}
\label{98da}f_{\beta _{2k}}\left( \Phi ^{\beta _{2k}}\right) \neq
Z_{\;\;\beta _{2k}}^{\beta _{2k-1}}\rho _{\beta _{2k-1}},
\end{equation}
\begin{equation}
\label{98c}g_{\beta _{2k+2}}\left( \Phi ^{\beta _{2k+2}}\right) \neq
A_{\beta _{2k+2}}^{\;\;\beta _{2k+3}}\gamma _{\beta _{2k+3}}\left( \Phi
^{\beta _{2k+2}}\right) ,
\end{equation}
for some functions $\rho _{\beta _{2k-1}}$ and $\gamma _{\beta _{2k+3}}$. On
account of (\ref{98da}--\ref{98c}) it is easy to show that the gauge
conditions (\ref{98a}) are irreducible even if $A_{\beta _k}^{\;\;\beta
_{k+1}}A_{\beta _{k+1}}^{\;\;\beta _{k+2}}\approx 0$. Thus, a possible class
of gauge-fixing fermions can be written as 
\begin{equation}
\label{99}\psi =\sum_{k=0}^b\bar \eta ^{\beta _{2k+1}}\chi _{\beta _{2k+1}},
\end{equation}
with $\chi _{\beta _{2k+1}}$ given in (\ref{98a}). Eliminating the
antifields from $S$ with the help of (\ref{99}), we deduce the gauge-fixed
action, $S_\psi $, in the standard manner. The gauge-fixing fermion (\ref{99}%
) involves (Dirac) $\delta $-functions from the gauge conditions. It is
understood that one can shift the gauge conditions by $B_{\beta _{2k+1}}$ in
order to reach some Gaussian average representations. Because the gauge
conditions are irreducible, the gauge-fixed action displays no residual
gauge invariances with respect to the non-minimal sector. Of course, one is
free to take any consistent irreducible gauge conditions instead of (\ref
{98a}). In conclusion, the path integral of the original reducible theory
quantized accordingly our irreducible procedure reads as 
\begin{equation}
\label{101}Z_\psi =\int D\Phi ^{A_0}D\eta ^{A_1}D\bar \eta
^{A_1}DB^{A_1}\exp iS_\psi .
\end{equation}
We remark once again that our procedure does not involve ghosts for ghosts,
i.e., (\ref{101}) contains only ghost number one ghost fields. This
completes the description of our irreducible treatment for reducible gauge
theories.

At this stage, we can emphasize in a clearer manner the role of the newly
added fields, $\Phi ^{\alpha _{2k}}$, with $k>0$. In our formalism these
fields play a double role, namely, (i) they implement the irreducibility
through the gauge transformations (\ref{57}--\ref{59}), and (ii) they are
involved with the irreducible gauge-fixing procedure. In this light, these
fields are more relevant than the corresponding non-minimal ones appearing
during the gauge-fixing process from the reducible case because, while the
newly introduced fields prevent the appearance of the reducibility, the
non-minimal fields (in the reducible situation) are mainly an effect of the
reducibility, and, consequently, are more passive.

\section{Example: the Freedman-Townsend model}

Let us apply the prior investigated irreducible approach in the case of the
Freedman-Townsend model. We start with the Lagrangian action of the
non-abelian Freedman-Townsend theory \cite{ft} 
\begin{equation}
\label{ft1}S_0^L\left[ B_{\mu \nu }^a,A_\mu ^a\right] =\frac 12\int
d^4x\left( -B_a^{\mu \nu }F_{\mu \nu }^a+A_\mu ^aA_a^\mu \right) , 
\end{equation}
where $B_{\mu \nu }^a$ stands for an antisymmetric tensor field, and the
field strength, $F_{\mu \nu }^a$, is defined by $F_{\mu \nu }^a=\partial
_\mu A_\nu ^a-\partial _\nu A_\mu ^a-f_{\;\;bc}^aA_\mu ^bA_\nu ^c$. Action (%
\ref{ft1}) is invariant under the first-stage on-shell reducible gauge
transformations 
\begin{equation}
\label{ft3}\delta _\epsilon B_{\mu \nu }^a=\varepsilon _{\mu \nu \lambda
\rho }\left( D^\lambda \right) _{\;\;b}^a\epsilon ^{\rho b},\;\delta
_\epsilon A_\mu ^a=0, 
\end{equation}
with $\left( D^\lambda \right) _{\;\;b}^a=\delta _{\;\;b}^a\partial ^\lambda
+f_{\;\;bc}^aA^{\lambda c}$. The field equations deriving from (\ref{ft1})
read as 
\begin{equation}
\label{ft5}\frac{\delta S_0^L}{\delta B_{\mu \nu }^a}\equiv -\frac
12F_a^{\mu \nu }=0,\;\frac{\delta S_0^L}{\delta A_a^\mu }\equiv A_\mu
^a+\left( D^\lambda \right) _{\;\;b}^aB_{\lambda \mu }^b=0. 
\end{equation}
The non-vanishing gauge generators of (\ref{ft3}), $\left( Z_{\mu \nu \rho
}\right) _{\;\;b}^a=\varepsilon _{\mu \nu \lambda \rho }\left( D^\lambda
\right) _{\;\;b}^a$, admit the first-order on-shell reducibility relations 
\begin{equation}
\label{red1}\left( Z_{\mu \nu \rho }\right) _{\;\;b}^a\left( Z^\rho \right)
_{\;\;c}^b=-\varepsilon _{\mu \nu \lambda \rho }f_{\;\;cd}^a\frac{\delta
S_0^L}{\delta B_{\lambda \rho d}}, 
\end{equation}
where the first-stage reducibility functions are expressed by $\left( Z^\rho
\right) _{\;\;c}^b=\left( D^\rho \right) _{\;\;c}^b$.

The equivalencies between the general background exposed above and the model
under consideration read as $\Phi ^{\alpha _0}\leftrightarrow B_{\mu \nu }^a$%
, $\epsilon ^{\alpha _1}\leftrightarrow \epsilon ^{\rho b}$, $Z_{\;\;\alpha
_1}^{\alpha _0}\leftrightarrow \varepsilon _{\mu \nu \lambda \rho }\left(
D^\lambda \right) _{\;\;b}^a$, $Z_{\;\;\alpha _2}^{\alpha _1}\leftrightarrow
\left( D^\rho \right) _{\;\;c}^b$, such that $\alpha _0\leftrightarrow
\left( a,\mu \nu \right) $, $\alpha _1\leftrightarrow \left( b,\rho \right) $%
, $\alpha _2\leftrightarrow c$. The fields $A_\mu ^a$ were omitted as their
gauge transformations identically vanish, hence they do not contribute to
the irreducible treatment. In agreement with our construction, we introduce
the fields $\Phi ^{\alpha _2}$, which in this case are some scalar fields $%
\varphi ^a$. We take $A_{\alpha _1}^{\;\;\alpha _2}$ to be $\left( D_\mu
\right) _{\;\;b}^a$, hence the concrete form of (\ref{57}) is 
\begin{equation}
\label{red2}\delta _\epsilon \varphi ^a=\left( D_\mu \right)
_{\;\;b}^a\epsilon ^{\mu b}. 
\end{equation}
It is easy to see that the new gauge transformations, namely, (\ref{ft3})
and (\ref{red2}) form a complete set, the gauge algebra remaining abelian.
This guarantees the possibility of an appropriate construction of the
longitudinal differential along the gauge orbits. With these elements at
hand, we pass to the derivation of the gauge-fixed action of the irreducible
system associated with the Freedman-Townsend model. The minimal ghost
spectrum contains the fermionic ghost number one ghosts $\eta ^{\alpha
_1}=\left( \eta ^{\mu b}\right) $, while the minimal antifield spectrum is
organized as $\Phi _{\alpha _0}^{*}=\left( B_a^{*\mu \nu },\varphi
_a^{*}\right) $ and $\eta _{\alpha _1}^{*}=\left( \eta _{\rho b}^{*}\right) $%
, the former antifields having antighost number one and Grassmann parity
one, while the latter possess antighost number two and Grassmann parity
zero. With the above spectra at hand, the concrete form of the minimal
solution to the master equation reads as 
\begin{equation}
\label{ft34}S_{\min }=S_0^L+\int d^4x\left( B_a^{*\mu \nu }\varepsilon _{\mu
\nu \lambda \rho }\left( D^\lambda \right) _{\;\;b}^a\eta ^{\rho b}+\varphi
_a^{*}\left( D_\mu \right) _{\;\;b}^a\eta ^{\mu b}\right) . 
\end{equation}
Next, we focus on the gauge-fixing process. We take the gauge conditions as
in (\ref{98a}), i.e., 
\begin{equation}
\label{gfc}\chi _{\rho b}\equiv -\frac 12\varepsilon _{\mu \nu \lambda \rho
}\left( D^\lambda \right) _{\;\;b}^cB_c^{\mu \nu }+\left( D_\rho \right)
_{\;\;b}^c\varphi _c=0. 
\end{equation}
The prior gauge conditions are irreducible and are enforced via a
non-minimal sector of the type $\left( \bar \eta _{\mu a},b^{\mu a}\right) $
plus the corresponding antifields. The ghost numbers ($gh$) and Grassmann
parities ($\epsilon $) of the non-minimal fields read as $gh\left( \bar \eta
_{\mu a}\right) =-1$, $\epsilon \left( \bar \eta _{\mu a}\right) =1$, $%
gh\left( b^{\mu a}\right) =0$, $\epsilon \left( b^{\mu a}\right) =0$. The
features of their antifields follow from the standard BRST rules. Thus, the
non-minimal solution to the master equation is expressed through $S=S_{\min
}+\int d^4x\,\bar \eta _{\mu a}^{*}b^{\mu a}$. Taking the gauge-fixing
fermion $\psi =\int d^4x\,\chi _{\rho b}\bar \eta ^{\rho b}$, and
eliminating as usually the antifields, we arrive at the gauge-fixed action 
\begin{eqnarray}\label{gf}
& &S_\psi =S_0^L+\int d^4x\left( -\frac 12\left( \left( D_{\left[ \lambda
\right. }\right) _{\;\;a}^c\bar \eta _{\left. \rho \right] c}\right) \left(
D^{\left[ \lambda \right. }\right) _{\;\;b}^a\eta ^{\left. \rho \right]
b}-\right. \nonumber \\ 
& &\left. \left( \left( D^\rho \right) _{\;\;a}^c\bar \eta _{\rho c}\right) 
\left( \left( D_\mu \right) _{\;\;b}^a\eta ^{\mu b}\right) +
\chi _{\rho b}b^{\rho b}\right) , 
\end{eqnarray}
with $\chi _{\rho b}$ given by (\ref{gfc}). The symbol $\left[ \lambda \rho
\right] $ signifies the antisymmetry with respect to the Lorentz indices
between brackets. We remark that the gauge-fixed action resulting from our
irreducible procedure is Lorentz covariant.

Let us investigate now an abelian version of Freedman-Townsend theory that
is second-stage reducible. We start with the action 
\begin{equation}
\label{ex1}S_0^L\left[ B_a^{\mu \nu \rho },A_\mu ^a\right] =\frac 12\int
d^5x\left( \frac 1{3!}\varepsilon _{\mu \nu \lambda \rho \sigma }B_a^{\mu
\nu \lambda }F^{\rho \sigma a}+A_\mu ^aA_a^\mu \right) ,
\end{equation}
where $B_a^{\mu \nu \rho }$ stand for an abelian three-form, while $F^{\rho
\sigma a}$ is now abelian, i.e., $F^{\rho \sigma a}=\partial ^{\left[ \rho
\right. }A^{\left. \sigma \right] a}$. The gauge transformations of (\ref
{ex1}) read as 
\begin{equation}
\label{ex2}\delta _\epsilon B_a^{\mu \nu \rho }=\partial _{}^{\left[ \mu
\right. }\epsilon _a^{\left. \nu \rho \right] },\;\delta _\epsilon A_\mu
^a=0,
\end{equation}
and are off-shell second-stage reducible, with the first, respectively,
second stage reducibility functions given by 
\begin{equation}
\label{ex3}Z_{\gamma b}^{\alpha \beta a}=\delta _b^{\;\;a}\partial
_{}^{\left[ \alpha \right. }\delta _\gamma ^{\left. \beta \right]
},\;Z_c^{\gamma b}=\delta _c^{\;\;b}\partial ^\gamma .
\end{equation}
Accordingly the general theory, we add the fields $B_a^\mu $ that play the
role of $\Phi ^{\alpha _2}$'s, the gauge parameters $\epsilon _a$, which are
the analogues of $\epsilon ^{\alpha _3}$, and take $A_{\alpha
_1}^{\;\;\alpha _2}$ to be the transposed of $Z_{\gamma b}^{\alpha \beta a}$%
. Therefore, the gauge transformations of $B_a^\mu $ take the form (see (\ref
{57})) 
\begin{equation}
\label{ex4}\delta _\epsilon B_a^\mu =2\partial _\nu \epsilon _a^{\nu \mu
}+\partial ^\mu \epsilon _a.
\end{equation}
The non-minimal solution to the master equation is expressed by%
\begin{eqnarray}\label{ex5}
& &S=S_0^L\left[ B_a^{\mu \nu \rho },A_\mu ^a\right] +\int d^5x\left( B_{\mu
\nu \rho }^{*a}\partial _{}^{\left[ \mu \right. }\eta _a^{\left. \nu \rho
\right] }+\right. \nonumber \\ 
& &\left. B_\mu ^{*a}\left( 2\partial _\nu \eta _a^{\nu \mu
}+\partial ^\mu \eta _a\right) +\bar \eta _a^{*\mu \nu }b_{\mu \nu }^a+\bar
\eta _a^{*}b^a\right) . 
\end{eqnarray}
In the last formula $\left( \eta _a^{\mu \nu },\eta _a\right) $ denote the
fermionic ghost number one ghosts, $\left( B_{\mu \nu \rho }^{*a},B_\mu
^{*a}\right) $ are the fermionic ghost number minus one antifields of the
corresponding fields, while the remaining fields form the non-minimal
sector. If we take some gauge conditions as in (\ref{98a}) by means of the
gauge-fixing fermion 
\begin{equation}
\label{ex6}K=\int d^5x\left( \bar \eta _{\mu \nu }^a\left( \partial _\rho
B_a^{\rho \mu \nu }+\frac 12\partial _{}^{\left[ \mu \right. }B_a^{\left.
\nu \right] }\right) +\bar \eta ^a\partial _\mu B_a^\mu \right) ,
\end{equation}
we arrive at the gauge-fixed action%
\begin{eqnarray}\label{ex7}
& &S_K=S_0^L\left[ B_a^{\mu \nu \rho },A_\mu ^a\right] +\int d^5x\left( \bar
\eta _{\mu \nu }^a\Box \eta _a^{\mu \nu }+\bar \eta ^a\Box \eta _a+\right. 
\nonumber \\
& &\left. b_{\mu \nu }^a\left( \partial _\rho B_a^{\rho \mu \nu
}+\frac 12\partial _{}^{\left[ \mu \right. }B_a^{\left. \nu \right] }\right)
+b^a\partial _\mu B_a^\mu \right) , 
\end{eqnarray}
where $\Box =\partial _\lambda \partial ^\lambda $. It is easy to see that
the gauge-fixed action (\ref{ex7}) has no residual gauge invariances.

Finally, we make the comparison between our approach and the reducible BRST
treatment in the case of the investigated models. The gauge-fixed actions
for the former, respectively, latter model within the reducible treatment
are given by 
\begin{eqnarray}\label{ex8}
& &S_{\psi ^{\prime }}^{\prime }=S_0^L\left[ B_{\mu \nu }^a,A_\mu ^a\right]
+\int d^4x\left( -\frac 12\left( \left( D_{\left[ \mu \right. }\right)
_{\;\;a}^c\bar \eta _{\left. \nu \right] c}\right) \left( D^{\left[ \mu
\right. }\right) _{\;\;b}^a\eta ^{\left. \nu \right] b}-\right. 
\nonumber \\
& &\left( \left( D_\mu \right) _{\;\;a}^c\bar \eta _c^\mu \right) \left( \left(
D^\nu \right) _{\;\;b}^a\eta _\nu ^b\right) -\left( \left( D^\mu \right)
_{\;\;a}^c\bar C_c\right) \left( D_\mu \right) _{\;\;b}^aC^b+ 
\nonumber \\
& &\frac 18\epsilon ^{\mu \nu \lambda \rho }f_{\;\;bc}^a\left( 
\left( D_{\left[
\mu \right. }\right) _{\;\;a}^d\bar \eta _{\left. \nu \right] d}\right)
\left( \left( D_{\left[ \lambda \right. }\right) _{\;\;e}^c\bar \eta
_{\left. \rho \right] }^e\right) C^b+\nonumber \\ 
& &\left. \left( -\frac 12\epsilon ^{\mu \nu \lambda \rho }\left(
D_\nu \right) _{\;\;a}^bB_{\lambda \rho b}+\left( D^\mu \right)
_{\;\;a}^b\bar \eta _b\right) b_\mu ^a\right) , 
\end{eqnarray}
\begin{eqnarray}\label{ex9}
& &S_{K^{\prime }}^{\prime }=S_0^L\left[ B_a^{\mu \nu \rho },A_\mu ^a\right]
+\int d^5x\left( \bar \eta _{\mu \nu }^a\Box \eta _a^{\mu \nu }+\bar \eta
^a\Box \eta _a+\right. \nonumber \\
& &b_{\mu \nu }^a\left( \partial _\rho B_a^{\rho \mu \nu }+\frac 12\partial
_{}^{\left[ \mu \right. }\bar \eta _a^{\left. \nu \right] }\right)
+b^a\partial _\mu \bar \eta _a^\mu -\partial _{\left[ \mu \right. }\bar \eta
_{\left. \nu \right] }^{\prime a}\partial _{}^{\left[ \mu \right.
}C_a^{\left. \nu \right] }-\nonumber \\ 
& &\left. \left( \partial _\mu \bar \eta ^{\prime a}\right) \left(
\partial ^\mu C_a\right) -\left( \partial ^\mu \bar \eta _\mu ^{\prime
a}\right) b_a^{\prime }+\left( \partial _\mu C_a^\mu \right) b^{\prime
\prime a}\right) . 
\end{eqnarray}
Apart from the spectra in the irreducible setting, in (\ref{ex8}) there
appear the additional variables $C^a$, that are the bosonic ghost number two
ghosts, plus the non-minimal fields $\left( \bar C_a,\bar \eta _a\right) $.
Along the same line, but with respect to the gauge-fixed action (\ref{ex9}),
the fields $C_a^\mu $ and $C_a$ stand for the ghost number two,
respectively, three ghosts, while $\left( \bar \eta _a^\mu ,\bar \eta _\mu
^{\prime a},\bar \eta ^{\prime a},b_a^{\prime },b^{\prime \prime a}\right) $
belong to the non-minimal sector. By performing the identifications 
\begin{equation}
\label{ex10}\bar \eta ^a\leftrightarrow \varphi ^a,
\end{equation}
respectively, 
\begin{equation}
\label{ex11}\bar \eta _a^\mu \leftrightarrow B_a^\mu ,
\end{equation}
between the variables involved with the gauge-fixed actions derived within
the irreducible and reducible approaches, the difference between the
gauge-fixed actions respectively corresponding to the two models are 
\begin{eqnarray}\label{ex12} 
& &S_{\psi ^{\prime }}^{\prime }-S_\psi =
\int d^4x\left( -\left( \left( D^\mu
\right) _{\;\;a}^c\bar C_c\right) \left( D_\mu \right) _{\;\;b}^aC^b+\right. 
\nonumber \\
& &\left. \frac 18\epsilon ^{\mu \nu \lambda \rho
}f_{\;\;bc}^a\left( \left( D_{\left[ \mu \right. }\right) _{\;\;a}^d\bar
\eta _{\left. \nu \right] d}\right) \left( \left( D_{\left[ \lambda \right.
}\right) _{\;\;e}^c\bar \eta _{\left. \rho \right] }^e\right) C^b\right) , 
\end{eqnarray}
\begin{eqnarray}\label{ex13}
& &S_{K^{\prime }}^{\prime }-S_K=
\int d^5x\left( -\partial _{\left[ \mu \right.
}\bar \eta _{\left. \nu \right] }^{\prime a}\partial _{}^{\left[ \mu \right.
}C_a^{\left. \nu \right] }-\right. \nonumber \\
& &\left. \left( \partial _\mu \bar \eta ^{\prime a}\right) \left(
\partial ^\mu C_a\right) -\left( \partial ^\mu \bar \eta _\mu ^{\prime
a}\right) b_a^{\prime }+\left( \partial _\mu C_a^\mu \right) b^{\prime
\prime a}\right) . 
\end{eqnarray}
We remark that the differences between the gauge-fixed actions are
proportional with the ghosts of ghost number greater than one, which are
some essential compounds of the reducible BRST quantization. Although
identified at the level of the gauge-fixed actions, the fields from (\ref
{ex10}--\ref{ex11}) play different roles within the two formalisms. More
precisely, the presence of the fields $\varphi ^a$ and $B_a^\mu $ prevents
the reducibility, while the $\bar \eta ^a$'s, respectively, $\bar \eta
_a^\mu $ represent an effect of the reducibility. In fact, the fields $%
\varphi ^a$ and $B_a^\mu $ are introduced in order to forbid the existence
of the zero modes. In consequence, all the ingredients connected with the
zero modes, e.g., the ghosts of ghosts or the non-minimal pyramid, are
discarded from the irreducible setting. In this light, we suggestively call
the fields $\varphi ^a$ and $B_a^\mu $ `antimodes'. This completes our
irreducible treatment.

\section{Conclusion}

To conclude with, in this paper we expose an alternative method of
quantizing reducible gauge theories without introducing ghosts of ghosts or
their antifields. The cornerstone of our approach is given by the derivation
of a Koszul-Tate complex underlying an irreducible gauge theory. As the
irreducible gauge system possesses the same physical observables like the
original reducible theory, it is legitimate to substitute the BRST
quantization of the initial reducible system by that of the irreducible one
from the point of view of the main equations underlying the BRST formalism.
Then, the general line of the antifield BRST quantization for the
irreducible theory is elucidated, some possible gauge conditions being
outlined. The general approach is finally exemplified in the case of the
Freedman-Townsend model.

\end{document}